\documentclass[twocolumn]{aastex63}
\usepackage{amsmath}

\usepackage{graphicx}
\usepackage{amsmath}
\usepackage{mathtools}
\usepackage{kantlipsum}
\usepackage{xparse}
\usepackage{enumerate}
\usepackage[version=3]{mhchem}
\usepackage{amsmath,array}
\usepackage[T1]{fontenc}
\usepackage{courier}
\usepackage{booktabs}

\usepackage{sidecap}
\usepackage{tabularx}

\submitjournal{ApJ}

\graphicspath{{./}{figures/}}

\begin{document}

\title{SPORK That Spectrum: Increasing Detection Significances from High-Resolution Exoplanet Spectroscopy with Novel Smoothing Algorithms}

\author[0000-0002-0470-0800]{Kaitlin C. Rasmussen}
\affil{Department of Astronomy and Astrophysics, University of Michigan, Ann Arbor, MI, 48109, USA}

\author[0000-0002-7704-0153]{Matteo Brogi}
\affil{Department of Physics, University of Warwick, Coventry CV4 7AL, UK}
\affil{INAF—Osservatorio Astrofisico di Torino, Via Osservatorio 20, I-10025, Pino Torinese, Italy}
\affil{Centre for Exoplanets and Habitability, University of Warwick, Gibbet Hill Road, Coventry CV4 7AL, UK}

\author[0000-0001-7340-5912]{Fahin Rahman}
\affil{Department of Astronomy and Astrophysics, University of Michigan, Ann Arbor, MI, 48109, USA}

\author[0000-0003-3963-9672]{Emily Rauscher}
\affil{Department of Astronomy and Astrophysics, University of Michigan, Ann Arbor, MI, 48109, USA}

\author[0000-0002-6980-052X]{Hayley Beltz}
\affil{Department of Astronomy and Astrophysics, University of Michigan, Ann Arbor, MI, 48109, USA}

\author[0000-0002-4863-8842]{Alexander P. Ji} 
\affiliation{Observatories of the Carnegie Institution for Science, 813 Santa Barbara St., Pasadena CA 91101}
\affiliation{Department of Astronomy \& Astrophysics, University of Chicago, 5640 S Ellis Avenue, Chicago, IL 60637, USA}
\affiliation{Kavli Institute for Cosmological Physics, University of Chicago, Chicago, IL 60637, USA}

\begin{abstract}
Spectroscopic studies of planets outside of our own solar system provide some of the most crucial information about their formation, evolution, and atmospheric properties. In ground-based spectroscopy, the process of extracting the planet’s signal from the stellar and telluric signal has proven to be the most difficult barrier to accurate atmospheric information. However, with novel normalization and smoothing methods, this barrier can be minimized and the detection significance dramatically increased over existing methods. In this paper, we take two examples of CRIRES emission spectroscopy taken of HD 209458 b and HD 179949 b and apply SPORK (SPectral cOntinuum Refinement for telluriKs) and iterative smoothing to boost the detection significance from 5.78 to 9.71 sigma and from 4.19 sigma to 5.90 sigma, respectively. These methods, which largely address systematic quirks introduced by imperfect detectors or reduction pipelines, can be employed in a wide variety of scenarios, from archival data sets to simulations of future spectrographs. 

\end{abstract}
\keywords{planets and satellites: atmospheres}

\section{Introduction}\label{sec:Introduction}

The high-resolution (R $\gtrsim$ 25,000) spectroscopic study of exoplanets, which has unlocked characterization of atmospheres at unprecedented scales, began in 1999 with \citet{Charbonneau1999}, who used optical time-series spectra of the hot Jupiter $\tau$ Bootis b to claim a low albedo for the planet. Later that year, the first direct detection of reflected light from the same planet was made by \citet{CollierCameron1999}, and in the coming decade, more attempts would be made \citep{CollierCameron2002, Leigh2003, Rodler2008}. A number of further attempts were performed to characterize the atomic and molecular make-up of $\tau$ Bootis b as well as other hot Jupiters known at the time \citep{Wiedemann2001, Drake2005}, but the first successful high-resolution measurements of an atomic species in an exoplanet atmosphere were \citet{Snellen2008} and \citet{Redfield2008}, who both measured optical Na lines via transmission spectroscopy in the planets HD 209458 b and HD 189733 b, respectively. Later, Snellen would be the first to utilize cross-correlation as a detection method \citet{Snellen2010}. High-resolution \textit{emission} spectroscopy was introduced by \citet{Mandell2011}, who used Keck’s NIRSPEC to dispute a low-resolution claim of water in the L-band of HD 189733 b \citep{Barnes2010}. Since then, the field has expanded considerably with the growing availability of high-resolution spectrographs, large space- and ground-based telescopes, and increasingly effective statistical analysis methods. An excellent summary of the methodology of high-resolution exoplanet spectroscopy is presented in \citet{Birkby2018}.

Since its conception, however, the study of exoplanet atmospheres has been hindered by the difficulty of removing the stellar and telluric components of the spectra, which make up the vast majority of the signal. Stellar lines, for the most part, do not vary over an observation. Tellurics, on the other hand, being a part of Earth’s ever-changing atmosphere, can vary strongly even from spectrum to spectrum depending on a number of factors such as water vapor density, season, cloud coverage, and zenith angle of the observation. Analysis of the exoplanet spectrum cannot begin until both of these components are removed.

Over the years, several measures have been adopted to remove stellar and telluric lines either separately or together. \citet{Snellen2008} as well as many later emission spectroscopy papers \citep{Brogi2012,Brogi2014,Brogi2016}, perform a relatively simple and fast removal by linearly fitting airmass trends and dividing out a ‘reference’ (median) spectrum. This method works because the position of the exoplanet spectrum \textit{moves} over pixels, whereas the position and depth of the stellar lines, and the position (but not the depth) of the tellurics remains static. The results of this method can then be directly cross-correlated with the atmospheric model.




Another method for telluric removal is Principal Component Analysis (PCA). PCA decomposes a set of vectors into a linear combination of eigenvectors and eigenvalues. In this case, the input vectors can be the individual spectra (i.e. PCA in the wavelength domain, see \citet{Lockwood2014,Piskorz2016,Piskorz_2017,Giacobbe2021}); or the individual spectral channels (PCA in the time domain, see \citet{deKok2013}). Regardless of the choice of domain, PCA identifies ``components'', i.e. trends that are in common mode between all the input vectors. The user can select the minimum set of these trends sufficient to reproduce the main variations in the data, and divide their linear combination out to essentially normalise the data. 

The program SYSREM \citep{Tamuz2005} is a PCA routine which can account for error bars which are not static. This is crucial because the error bars on each point in a spectrum are correlated with where the point falls on the spectrograph; i.e., points which fall in the center pixels of the CCD have a higher SNR than pixels at the edge. SYSREM disassembles time-series spectra into principal components and removes them to an order chosen by the user. This method has been used successfully in emission spectra by \citet{Birkby2013,Nugroho2017,Alonso-Floriano2019,Sanchez-Lopez2019}, and many others.

In this paper, we focus on the former of these methods: airmass detrending. Airmass detrending is simple and fast, and unlike PCA and SYSREM, has a high degree of user control and flexibility, making it an ideal testing ground for our new methods.


\section{Data \& Models}\label{sec:DataModels}

We take two examples of hot Jupiters: the well-known HD 209458 b, and the lesser-studied HD 179949 b, and detect CO in the former, and CO and H$_{2}$O in the latter.

\textbf{HD 209458 b:} HD 209458 b is a canonical example of a hot Jupiter, with a mass and radius of 0.67 M$_{Jupiter}$ and 1.38 R$_{Jupiter}$, respectively, and a period of 3.52 days \citep{Southworth2010}. Spitzer phase curves of thermal emission from the planet show a dayside brightness temperature of 1499 $\pm$ 15 K and 972 $\pm$ 44 K on the nightside \citep{Zellem2014}. Many species have been detected in its atmosphere, including He \citep{Alonso-Floriano2019}, water vapor \citep{Sanchez-Lopez2019}, and several detections of CO \citep{Gandhi2019, Beltz2021, Giacobbe2021}, which we focus on in this work. 

We use simulated spectra generated from a 3D General Circulation Model (GCM) of HD 209458 b, post-processed with a line-by-line radiative transfer routine that only includes CO as an opacity source. Details on the GCM can be found in \citet{Beltz2021}, and more information about the post-processing routine used can be found in \citet{Zhang2017}. Due to the strong influence the underlying chemistry assumptions on the detection significance of the models containing water in \citet{Beltz2021}, we chose to use the CO-only model spectra  which was more robust in significance under different atmospheric chemistry assumptions in this work.

The spectroscopic data was originally published in \citet{Schwarz2015}, and detailed information about the observations (CRIRES; 2.285--2.348 micron, R $\sim$ 100,000) can be found there. The spectra were optimally extracted using the ESO pipeline \citep{Freudling2013} and wavelength-calibrated using the known positions of telluric lines. The data was also used in \citet{Beltz2021} in order to test the efficacy of 3D models vs. 1D models in emission spectra.

\textbf{HD 179949 b:} HD 179949 b is a non-transiting hot Jupiter similar to HD 209458 b, with a mass of 0.92 M$_{Jupiter}$  \citep{Wang2011}, a period of 3.09 days \citep{Wittenmyer2007}, and a zero-albedo equilibrium temperature of about 1600 K. A Spitzer phase curve of this planet implied fairly inefficient transport of heat to the planet's night side \citep{Cowan2007}. H2O and CO have been detected in its atmosphere by \citet{Brogi2014} and \citet{Webb2020}.  

The atmospheric model used to cross-correlate against the spectra of HD 179949 was chosen from a large grid of models described in \citet{Brogi2014}. A model with VMR(CO) = VMR(H2O) = 10$^{-4.5}$, VMR(CH4) = 10$^{-9.5}$, and a steep lapse rate of $\frac{dT}{dlog(\rho)}$ $\sim$ 330 K per pressure decade was the best fit to the data in that work, and thus is our choice for this analysis.

The spectroscopic data, taken at the same spectrograph, resolution, and wavelength range as HD 209458 b (CRIRES; 2.285--2.348 micron, R $\sim$ 100,000) was originally published in \citet{Brogi2014}, which thoroughly details its observation and calibration. The spectra we use in this analysis have been reduced and wavelength-corrected using the same methods as HD 209458 b. As in \citet{Beltz2021}, the fourth detector has been discarded due to known odd-even pixel effects.

\section{Methods}\label{sec:Methods}

\subsection{SPORK and Iterative Smoothing}


In a typical high-resolution observation of an exoplanet, the information is extracted by cross correlating a normalized model spectrum against the telluric-removed spectra. Any deviations from the continuum on either side reduce this significance. Here we show that correcting even slight deviations in the spectrum normalization can increase the significance of detections. It should be noted that regardless of the technique used to isolate the exoplanet spectrum---even if it is not airmass detrending---as long as it is within the framework of high-resolution cross-correlation spectroscopy, SPORK in particular can be utilized to improve detection significance. 

SPORK (SPectral cOntinuum Refinement for telluriKs)\footnote{https://github.com/fahin1/SPORK-Iterative-Smoothing} is a spectrum normalization routine adapted from the stellar abundance determination software Spectroscopy Made Hard\footnote{The most recent version of SMH, Spectroscopy Made Hard(er), can be found at https://github.com/andycasey/smhr. SMH was first described in \citealt{Casey2014}}. The iterative smoothing function is based on the python scipy package \texttt{interpolate}. 

\subsubsection{Better Spectrum Normalization with SPORK}
SPORK is a normalization tool designed specifically to locate a continuum even in the presence of many large dips, such as the absorption features one encounters in stellar spectra. The tool iteratively fits a univariate natural cubic spline and identifies outlier pixels to be sigma-clipped. Each order of an echelle spectrum has highly varying signal-to-noise as a function of wavelength, so the spectrum uncertainties reported by the pipeline are used to perform sigma clipping (rescaling by the standard deviation of the error-normalized deviations). Spline knots can be placed arbitrarily, but we use $N/2 - 1$ evenly spaced knots for each order (where $N$ is length of the wavelength array). The lower sigma clipping threshold is set to 1.0, as points even a little below the continuum belong to stellar or telluric lines and should be ignored. The upper sigma clipping threshold is set to 5.0, as features which rise sharply above a stellar spectrum are typically cosmic rays which should also be ignored. 

\begin{figure}
\includegraphics[scale=0.43]{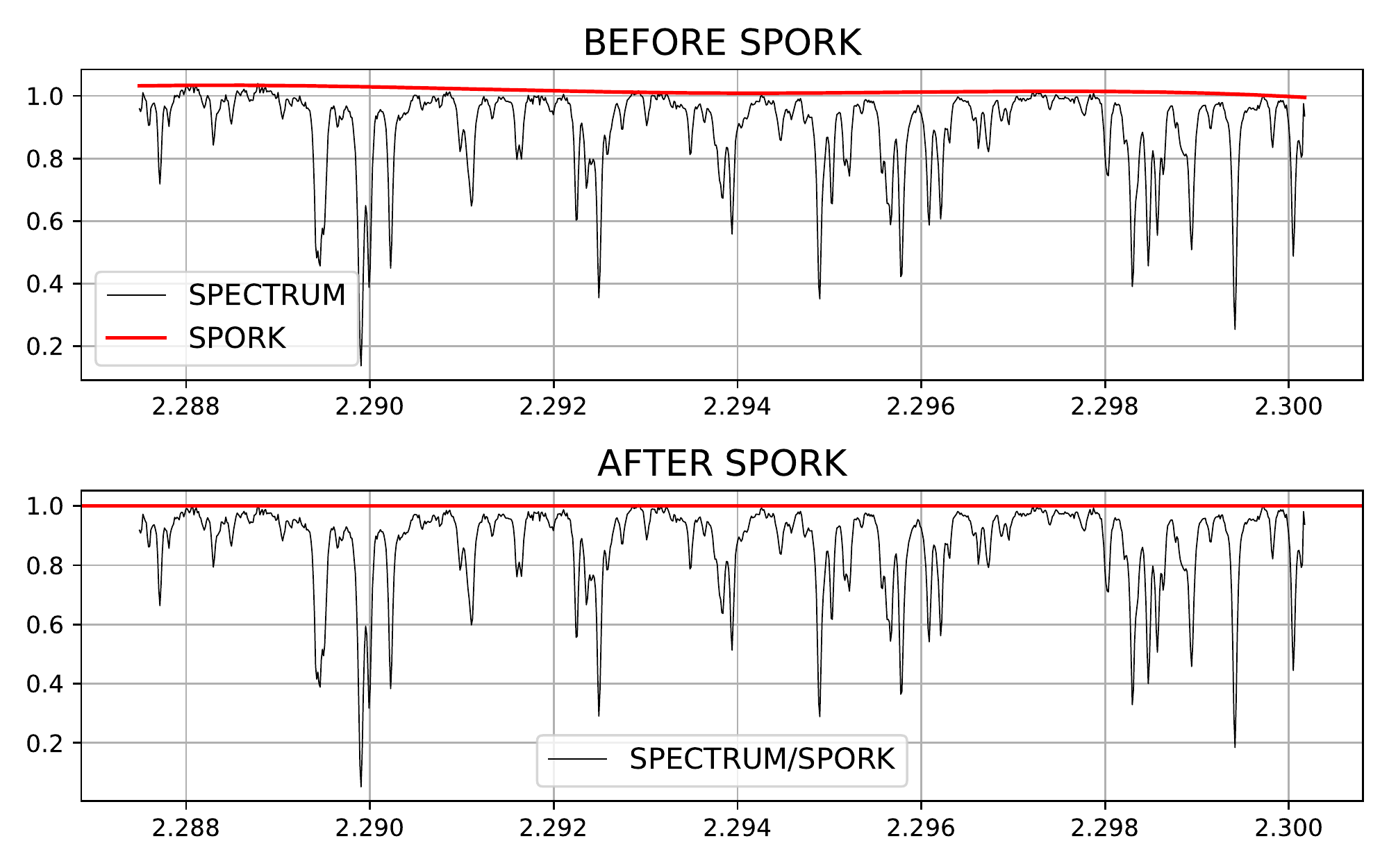}
\caption{SPORK applied to CRIRES spectra of HD 209458 b from the ESO reduction pipeline prior to telluric removal. Although the blaze function has been removed within the pipeline, residual wiggles in the continuum can be seen, especially between 2.288 and 2.292 microns, where the un-normalized continuum rises $\sim$ 5\% above the ``true'' continuum. If left uncorrected, these wiggles will persist throughout the telluric removal process and hinder the cross-correlation of the data against a perfectly flat model spectrum.} 
\label{fig:small_norm}
\end{figure}

\begin{figure}
\includegraphics[scale=0.42]{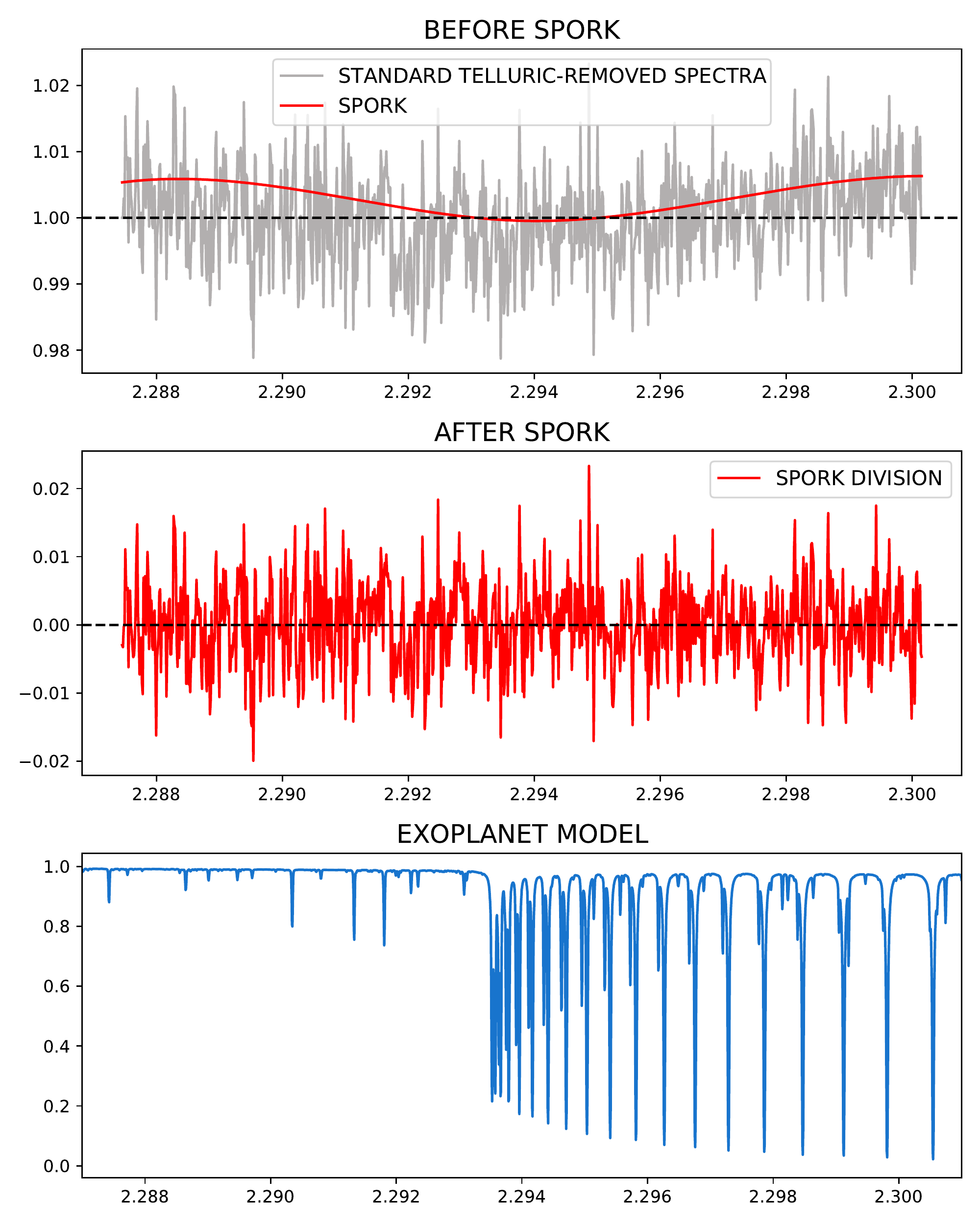}
\caption{SPORK applied to CRIRES spectra of HD 179949 b from the ESO reduction pipeline after the standard telluric removal process, described in Section \ref{sec:Methods}, has been performed. (Top):A telluric-removed spectrum (grey) displays a strong offset from the continuum, and is fit with SPORK (red). (Middle): the SPORK fit is divided out, leaving behind a flat telluric-removed spectrum. (Bottom): The normalized model exoplanet spectrum which corresponds to the telluric-removed spectral region is shown. Systematic offsets in the normalization of the HD 179949 data cannot be removed by performing SPORK on the individual spectra, as they persist in the median spectrum as well. In this case, it is better to fit the offset spectra to the offset median spectrum, and then SPORK residual. The flattened residual will perform better in the cross-correlation routine against the perfectly normalized exoplanet spectrum.} 
\label{fig:residual_norm}
\end{figure}

An example of one usage of SPORK, where the algorithm is applied before telluric removal, is shown in Fig \ref{fig:small_norm}. In this particular case, each spectrum shows slight deviations from the continuum which are averaged out, and thus SPORK should be applied \textit{before} telluric removal. However, this is not always the case; sometimes deviations in the spectra are systematic across the entire set. In this instance, it is more appropriate to apply the algorithm \textit{after} tellurics have been removed, such as in Fig \ref{fig:residual_norm}.

\subsubsection{Better Preservation of the Exoplanet Signal with Iterative Smoothing} Airmass detrending, the process by which stationary features of a spectrum are removed, is not a perfect process. Often, systematic noise is preserved as well, muddying the buried exoplanet spectrum. One way to circumvent this is to apply a small degree of smoothing to the median spectrum to de-noise it.   

\begin{figure}
\includegraphics[scale=.65]{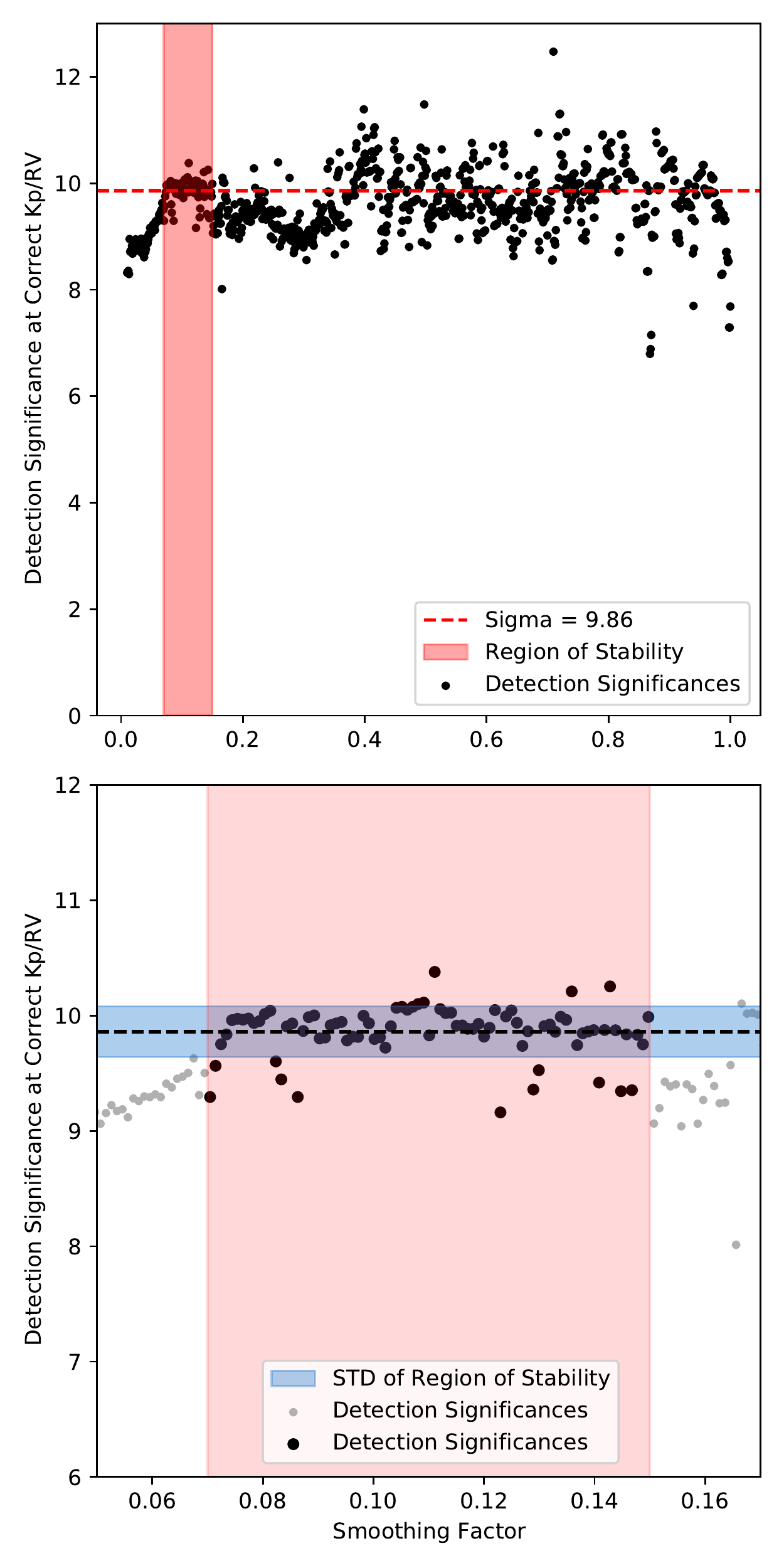}
\caption{Detection significance vs. smoothing factor for HD 209458 b. (Top): The full smoothing factor range from 0 to 1 is shown, with the region of stability highlighted in red. This region is chosen to avoid over-smoothing of the spectrum (see Fig \ref{fig:IS}, blue line), which begins to occur at 0.40 and produces high, but unstable detection significances. (Bottom): The region of stability is highlighted in blue. The values here have a mean of 9.86 and a standard deviation (STD) of 0.22. In our analysis we choose a smoothing factor of 0.12 (seen in Fig \ref{fig:IS}) which produces a detection significance of 9.82.} 
\label{fig:smoothing}
\end{figure}

In order to determine the best factor of smoothing, we first set the smoothing factor to zero. The python module scipy.interpolate (specifically, \texttt{splrep} and \texttt{splev}) is used to construct a spline fit to the median spectrum. We then apply the smoothing factor to the spline fit (in the first iteration, this will not affect the median spectrum). This way, instead of performing a second-order polynomial fit (polyfitting) of each individual spectrum to the median spectrum, each one is polyfit to the smoothed median spectrum instead. The temporal polyfitting is performed, and the entire model-injection cross-correlation routine used to detect the planet's signal in the data (details can be found in \citet{Beltz2021}) is run at the literature K$_{\rm P}$ and RV-rest position to determine the significance of the detection. We then increase the smoothing factor by 0.001, and the entire telluric removal and cross-correlation routine is performed again. This process is repeated 1,000 times with the smoothing factor ranging from 0 to 1. In this method, the smoothing factor function must reach stability (Fig~\ref{fig:smoothing}; the region of stability will vary from instrument to instrument---here we define it as standard deviation (STD) $\lesssim$ 0.25), at which point any factor in the stable region can be chosen as the set factor.


\begin{figure*}
\includegraphics[scale=.6]{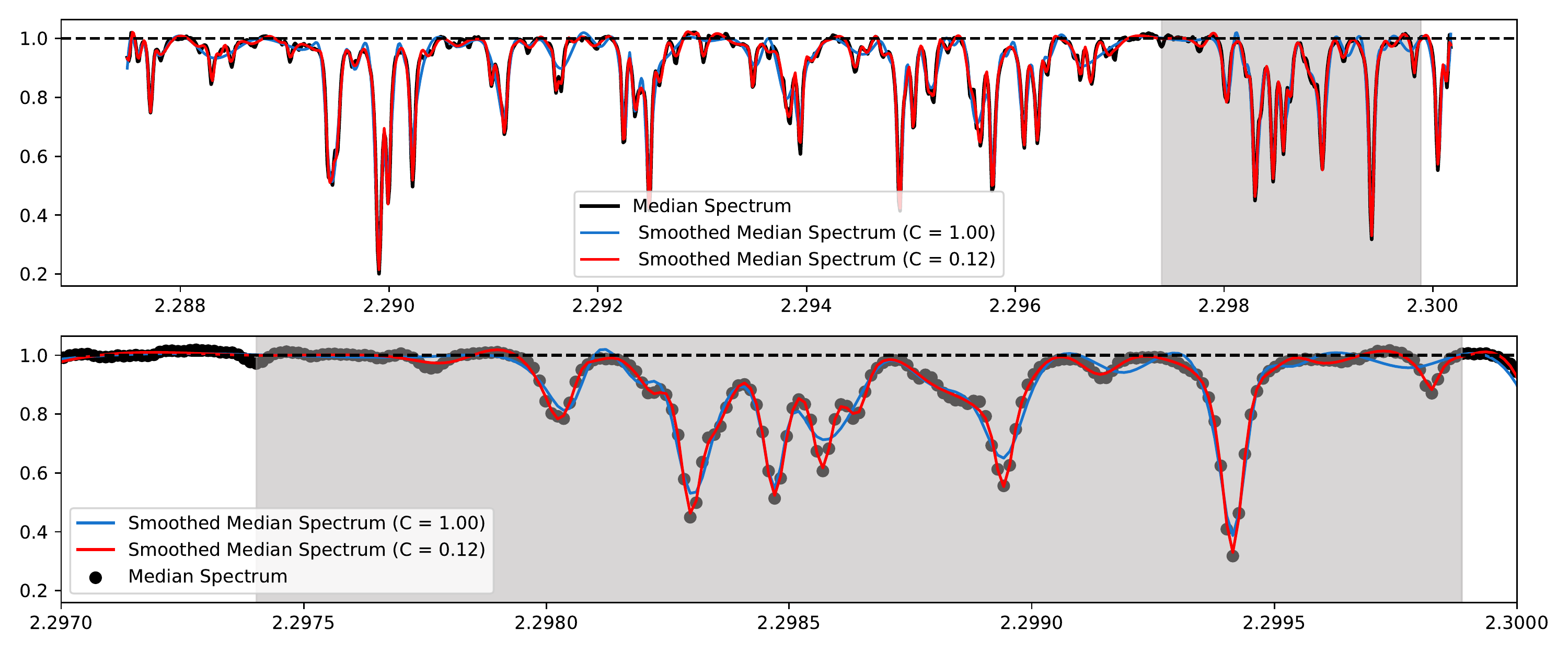}
\caption{A median CRIRES spectrum of HD 209458 b and its optimally smoothed version. In iterative smoothing, the smoothing factor C in the python function \texttt{splrep} is increased by 0.001 over a range of 0 to 1, and a value is chosen from the region of stability to be the final smoothing factor. The red line represents the optimal smoothing factor, while the blue line represents an overfit.} 
\label{fig:IS}
\end{figure*}

\subsection{Telluric Removal}

The method we follow to achieve our results, airmass detrending, is detailed in \citet{Brogi2019} and closely follows \citet{Beltz2021}, but will be recounted here. Airmass detrending is optimized for dayside observations, where the planet's disk is fully illuminated and its change in radial velocity is the fastest as it shifts from positive to negative around secondary eclipse. This rapid change in velocity means that the planet's signal is actually shifting across pixels on the detector, and thus when the median (stationary) elements of the time-series spectra are removed, the planet signal will be left behind. 

\subsubsection{Standard Telluric Removal:} First, a median spectrum is constructed for each time series. In the wavelength domain, each spectrum is fit against the median spectrum and a second-order polynomial is fit between the two, then divided out. This process is repeated in the time domain to account for temporal spectrum-to-spectrum changes in telluric abundances.

\subsubsection{Telluric Removal with SPORK:} As in the standard routine, outermost pixels are masked and the median spectrum is constructed. However, before a polynomial fit is derived, the spectrum undergoes a round of SPORK normalization which removes any residual wiggles left over from the reduction pipeline. After the continuum is located and the SPORK array is divided out, the wavelength and time polyfitting is carried out as usual.

\subsubsection{Telluric Removal with SPORK and Iterative Smoothing:} The outermost pixels are masked and the median spectrum is constructed. One round of SPORK is applied to remove residual continuum wiggles. Then the median spectrum is subjected to an iterative degree of smoothing, and a smoothing factor is chosen from the region of stability. An example of the smoothed median spectrum using a smoothing factor from the region of stability as well as one which demonstrates an ``overfit'' is shown in Fig~\ref{fig:IS}.

\subsubsection{Usage}

Any spectra, from any spectrograph and using any telluric removal method should be first visually inspected to determine whether deviations from the continuum are occurring spectrum-by-spectrum or whether the entire set deviates together. If the former, SPORK should be applied \textit{before} telluric removal; if the latter, it should be applied \textit{after}. For particularly messy spectra, SPORK may be used both before and after. Iterative smoothing works solely within the framework of airmass detrending, and should be tested as such. 


\section{Results}\label{sec:Results}

\begin{figure*}
\includegraphics[trim={4cm 0 0 0 },scale=.53]{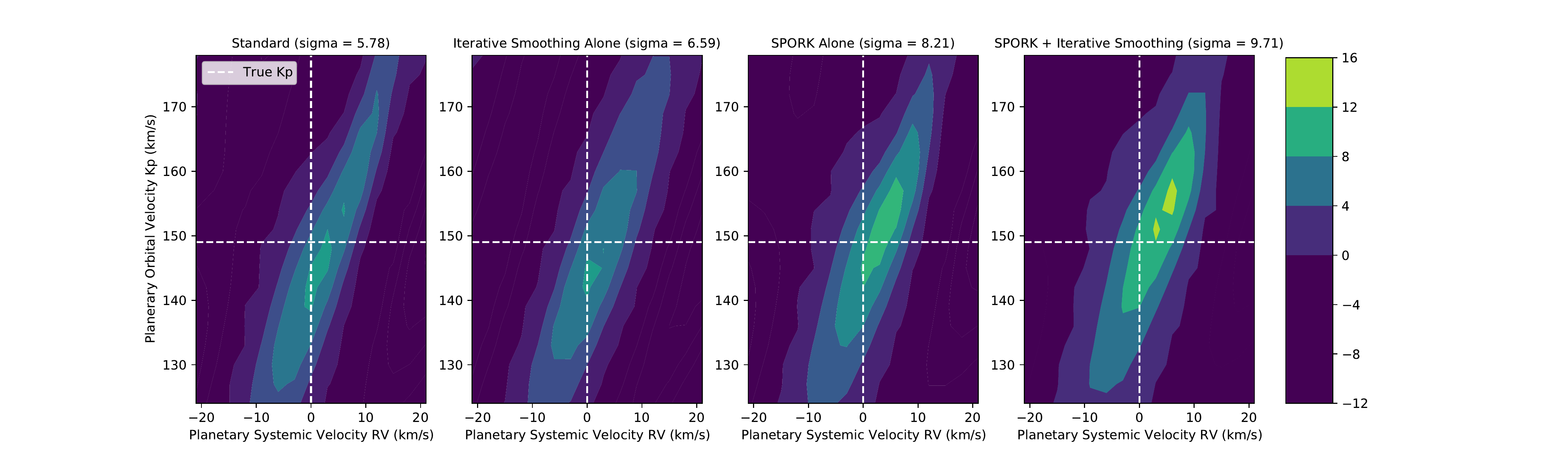}
\caption{Results of cross-correlation on telluric-removed spectra. (Left) The standard telluric routine leads to a detection of $\sigma$ = 5.78 at RV = 0, K$_{\rm P} = 149$, as is reported in \citet{Beltz2021}. (Middle) When SPORK is applied before the telluric removal process, the significance of the detection increases to $\sigma$ = 8.21. (Right) When SPORK is applied, then iterative smoothing is run, this factor increases to $\sigma$ = 9.71.} 
\label{fig:HD209458_result}
\end{figure*}

\begin{figure*}
\includegraphics[trim={4cm 0 0 0 },scale=.53]{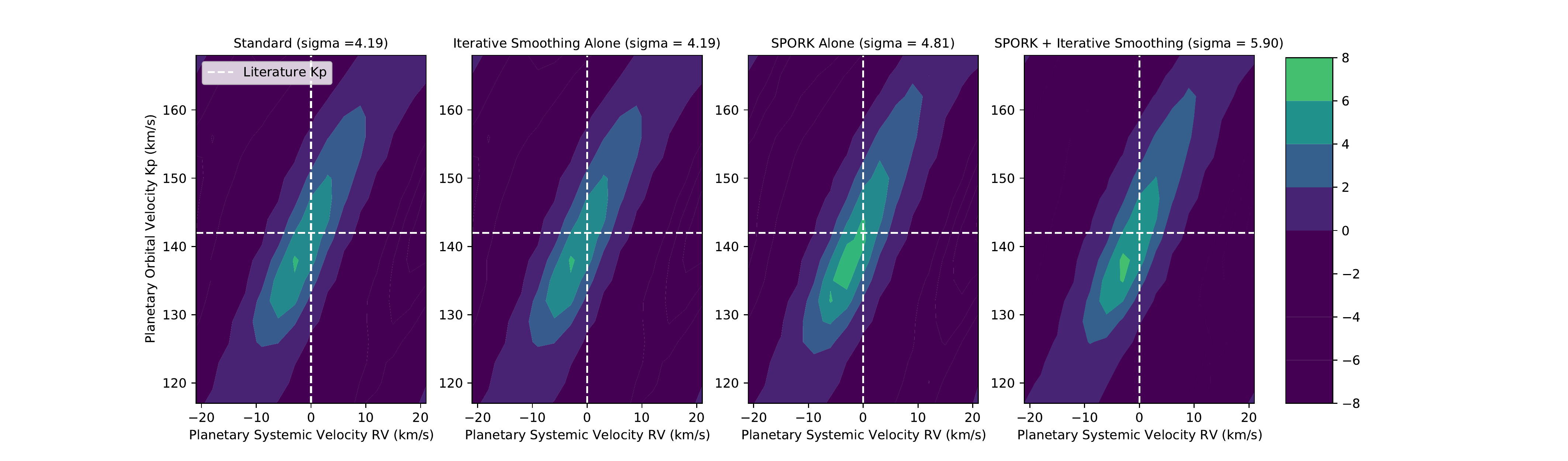}
\caption{Results of cross-correlation on telluric-removed spectra. (Left) The standard telluric routine leads to a detection of $\sigma$ = 4.19, similar to what is reported in \citet{Brogi2014}. (Middle Left): The inclusion of iterative smoothing on its own, in this case, does not result in any significant change to the detection significance. (Middle Right): When SPORK is applied after the telluric removal process, the significance of the detection increases to $\sigma$ = 4.81. (Right) When SPORK is applied, then iterative smoothing is run, this factor increases to $\sigma$ = 5.90.} 
\label{fig:HD179949_result}
\end{figure*}

The implementation of SPORK alone results in detection significance increases of 2.43 and 0.62 sigma at RV = 0 and the literature K$_{\rm P}$ for HD209458 b and HD 179949 b, respectively. The addition of iterative smoothing results in \textit{further} detection significance increases of 1.50 and 1.09. In total, we find that applying these new methods raised the detection significance of HD 209458 b and HD 179949 b from 5.78 $\sigma$ to 9.71 $\sigma$ and from 4.19 $\sigma$ to 5.90 $\sigma$, respectively. These results can be seen clearly in Figs. \ref{fig:HD209458_result} and \ref{fig:HD179949_result}. 

In the case of HD 209458 b, the application of SPORK to the spectra which come out of the ESO reduction pipeline has the stronger effect of the two methods. Not only does the signal increase dramatically, but the location of the ``smear'' of significance becomes more reflective of the literature K$_{\rm P}$ value. Applying an additional round of post-telluric-removal SPORK to the HD 209458 b analysis did not result in any further increase in signal.  For HD 179949 b, the location of the ``smear'' of significance does not change---this is perhaps due to the different SPORK treatments between the two planets, as well as the fact that HD 209458 b's atmosphere model was generated with a 3D GCM, while HD 179949 b's model was not.


This very strong increase in HD 209458's detection significance is likely due to the simple improvement of the cross-correlation. Because the planet's signal is weaker than the noise that surrounds it, it is crucial that minute shifts in the continuum are minimized, as they will create unwanted noise in the CCF.  

The effect of iterative smoothing is, in both cases, largely to keep the location of the detection in place, and to strengthen it. This is simply a manner of de-noising the SPORK-normalized spectrum. By smoothing over subtle systematic pixel-to-pixel noise, the exoplanet signal is revealed. This method, combined with SPORK, is clearly a powerful method of noise removal.

\subsection{Standard SNR and t-Tests}

\begin{figure*}
\includegraphics[trim={0cm 0 0 0 },scale=.68]{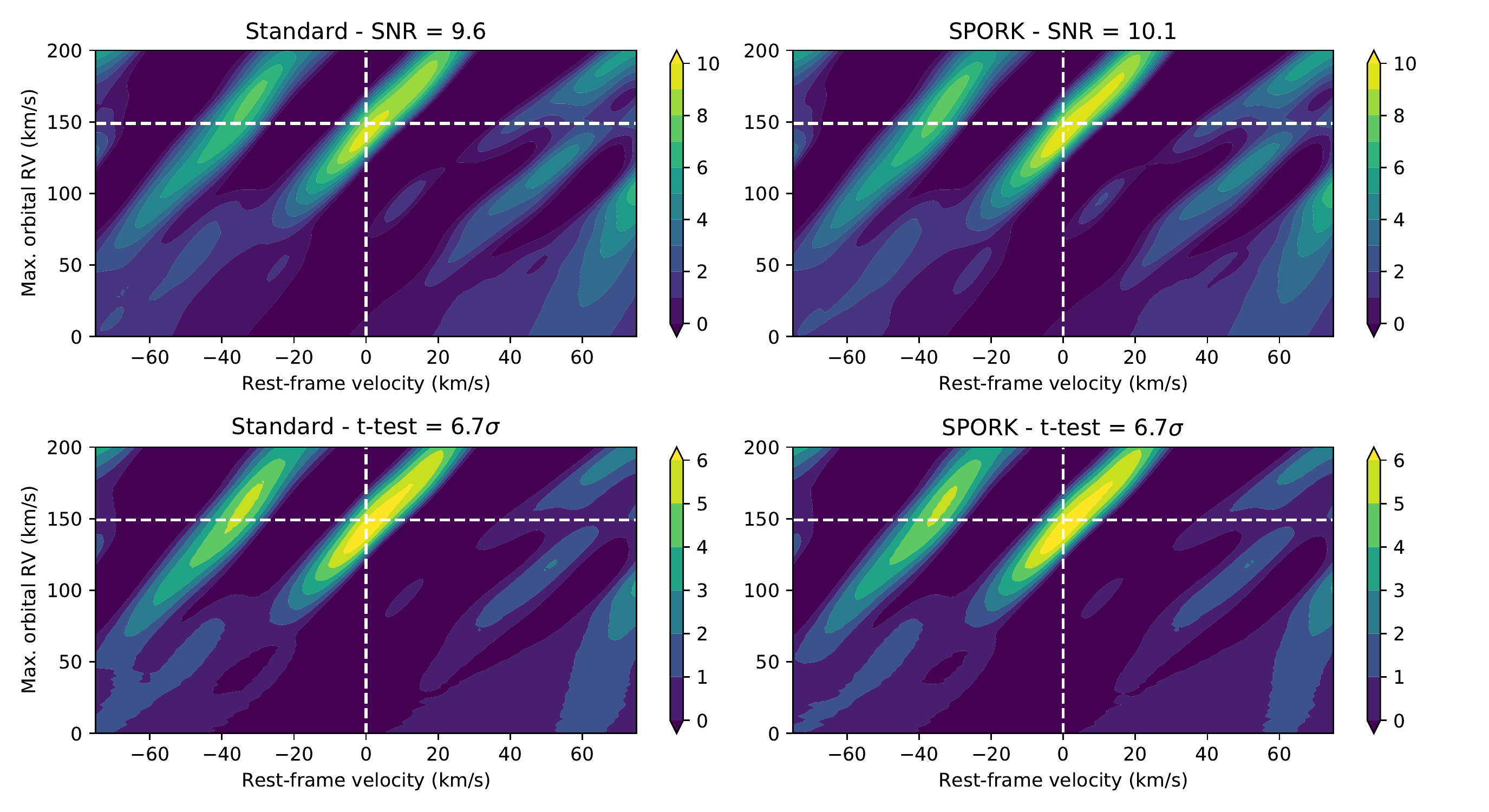}
\caption{Results of SNR and t-tests on spectral with and without SPORK/iterative smoothing performed for HD 209458 b. Top panels: the implementation of SPORK/iterative smoothing in SNR tests result in a 0.5 planetary SNR increase. Bottom panel: in the t-test application, the implementation of SPORK/iterative smoothing does not have an impact.} 
\label{fig:HD209458_SNR+ttest}
\end{figure*}

\begin{figure*}
\includegraphics[trim={0cm 0 0 0 },scale=.68]{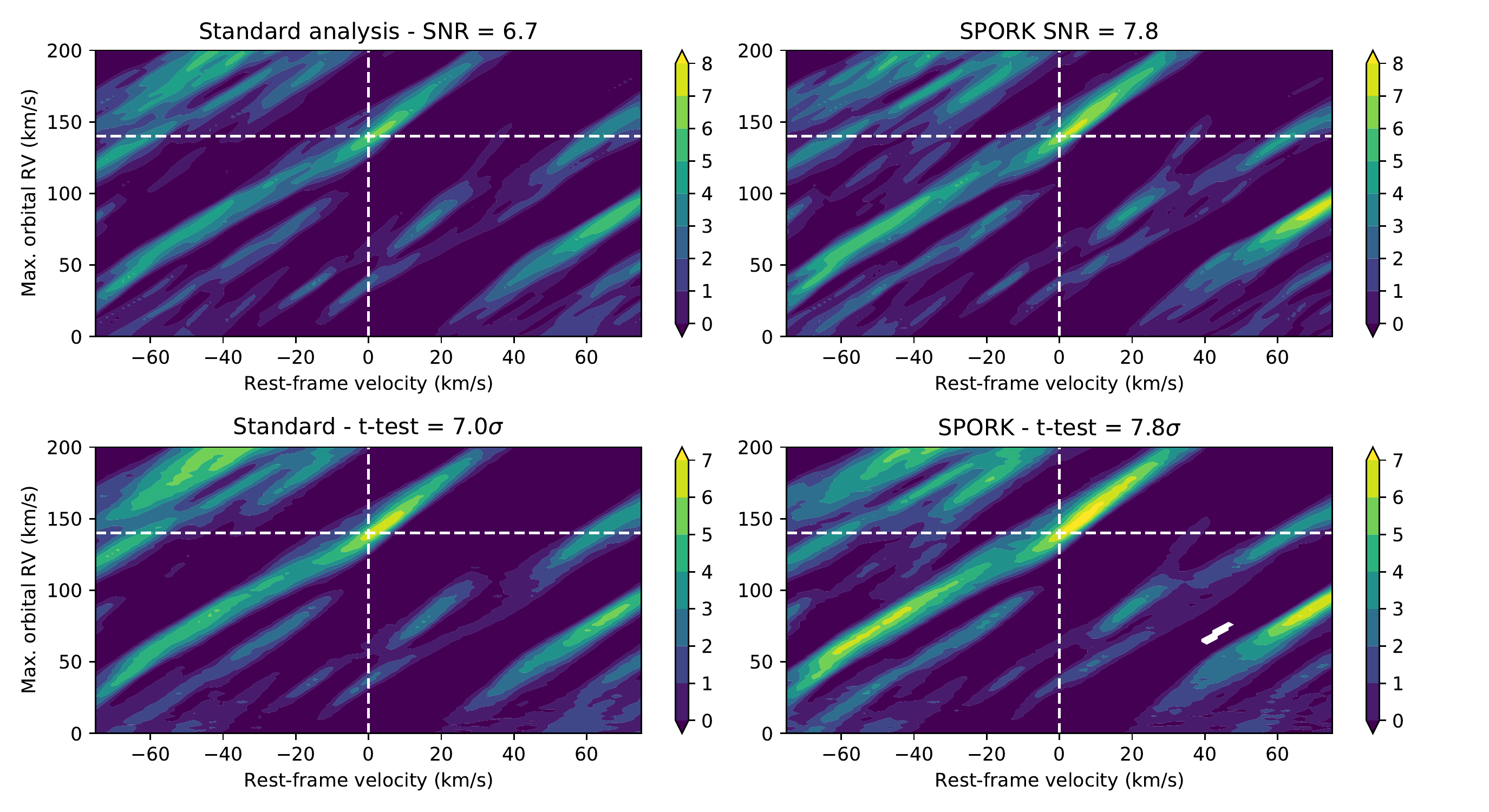}
\caption{Results of SNR and t-tests on spectra with and without SPORK/iterative smoothing performed for HD 179949 b. Top panels: the implementation of SPORK/iterative smoothing in SNR tests result in a 1.1 planetary SNR increase. Bottom panel: in the t-test application, the implementation of SPORK/iterative smoothing results in a 0.8-sigma increase.} 
\label{fig:HD179949_SNR+ttest}
\end{figure*}

We test SPORK and iterative smoothing in two different significance determination frameworks: SNR calculation and t-test. For HD 209458 b and HD 179949 b, the measured planetary SNR increases by 0.5 and 1.1, respectively (Fig~\ref{fig:HD209458_SNR+ttest}). When t-tests are performed, the significance either does not change (as in the case of HD 209458 b) or increases by 0.8 sigma (Fig~\ref{fig:HD179949_SNR+ttest}). 


\subsection{Discussion and Future Applications}

With the advent of large ground- and space-based observatories, and the coming era of terrestrial-sized exoplanet observations, SPORK and iterative smoothing will be essential to detecting and quantifying ever smaller exoplanet signals. They are simple and highly accessible techniques which can be applied to any set of spectra from any instrument in any wavelength range. Every spectrograph/reduction pipeline combination has systematic quirks that must be corrected for, and SPORK, in particular, can be applied not only to spectra that have already been reduced and normalized but even to spectra from which the blaze function has not been removed. For exoplanet science, this method can be utilized at any point before or during the telluric removal process, whether that process is airmass detrending, SYSREM, or model fitting.

The flexible SPORK can also be used to smooth, detrend, and normalize non-spectra data sets. In its current state, the knot spacing and the upper/lower sigma values are optimized for absorption spectra, but can also be customized for other one-dimensional data such as light curves. SPORK's one drawback is that it is not a fast algorithm, and thus significant overheads are expected when analysing large sets of data (modern spectrographs can produce datasets as big as 10$^7$ data-points) or comparing them to large grids of models (10$^5$-10$^6$), including the future integration into Bayesian retrievals.

Iterative smoothing is best applied within the airmass detrending framework. As it requires the running of the full cross-correlation code for many values, it is also quite slow (in its current iteration, 4--6 hours are required for 1000 smoothing factors---although it is only run one time) and should be used on a case-by-case basis. Future investigations, including efforts to parallelize this method, could yield lower computing times. 

One caveat of the current implementation of the analysis is that we are optimising the smoothing parameters by maximising the detection with a selected model. This could potentially lead to a model-dependent optimisation. Therefore, we recommend to first run the classic telluric removal without smoothing, and select a reasonable family of models that leads to a detection of the planet's atmosphere. The smoothing can then be optimised by using this sub-set of models, which would avoid cases in which a non-detection could be potentially overly-optimised to produce a detection.

Another important caveat is that in this work we explore the use of cross correlation techniques to ``detect'' a species, i.e. to maximise the information coming from matching the position and depth of a set of spectral lines. This application is crucial, e.g., to compile the chemical inventory of exoplanets, even when additional properties of the atmosphere are unknown. Recently, high-resolution spectroscopy has been extended to extract information about abundances and temperature, e.g. in \citet{Brogi2019}. In this latter case, the unavoidable alteration of the ``true'' exoplanet spectrum via the application of telluric removal needs to be thoroughly simulated and replicated on each model to avoid biases. Such level of detail is arguably beyond the scope of this work, and we defer to a follow-up study on the exploration of the use of SPORK within Bayesian retrievals of exoplanet atmospheres.


\acknowledgements
This research was supported by a grant from the Heising-Simons Foundation. KCR would like to thank Andy Casey for his work in developing the software Spectroscopy Made Hard(er) from which SPORK is derived.

\bibliography{bib}{}
\bibliographystyle{aasjournal}


\end{document}